\documentclass[twocolumn]{webofc}

\usepackage[varg]{txfonts}   
\begin{document}
\title{The Radio Detector of the Pierre Auger Observatory -- status and expected performance}
%
%

\author{\firstname{Tim} \lastname{Huege}\inst{1,2}\fnsep\thanks{\email{tim.huege@kit.edu}} for
        \lastname{the Pierre Auger Collaboration}\inst{3}\fnsep\thanks{\email{spokespersons@auger.org}, full author list available at https://www.auger.org/archive/authors\_2022\_10.html}
}

\institute{Karlsruhe Institute of Technology, Institute for Astroparticle Physics (IAP), Karlsruhe, Germany 
\and
           Vrije Universiteit Brussel, Astrophysical Institue, Brussels, Belgium
\and
           Observatorio Pierre Auger, Avenida San Mart{\'\i}n Norte 304, Malarg\"ue, Argentina
          }

\abstract{%
  As part of the ongoing AugerPrime upgrade of the Pierre Auger Observatory, we are deploying short aperiodic loaded loop antennas measuring radio signals from extensive air showers in the $30-80$\,MHz band on each of the 1,660 surface detector stations. This new Radio Detector of the Observatory allows us to measure the energy in the electromagnetic cascade of inclined air showers with zenith angles larger than $\sim 65^\circ$. The water-Cherenkov detectors, in turn, perform a virtually pure measurement of the muon component of inclined air showers. The combination of both thus extends the mass-composition sensitivity of the upgraded Observatory to high zenith angles and therefore enlarges the sky coverage of mass-sensitive measurements at the highest energies while at the same time allowing us to cross-check the performance of the established detectors with an additional measurement technique. In this contribution, we outline the concept and design of the Radio Detector, report on its current status and initial results from the first deployed stations, and illustrate its expected performance with a detailed, end-to-end simulation study.}

\newcommand{\Rmu}{\ensuremath{R_\mu}\xspace}
\newcommand{\Eem}{\ensuremath{E_\mathrm{em}}\xspace}
\newcommand{\Ecr}{\ensuremath{E_\mathrm{CR}}\xspace}
\newcommand{\mean}[1]{\ensuremath{\langle #1 \rangle}\xspace}

\maketitle
\section{Introduction}

The Pierre Auger Observatory is currently undergoing the AugerPrime upgrade \cite{Castellina:2019irv}. In addition to equipping every water-Cherenkov detector (WCD) with a surface scintillator detector (SSD) \cite{BeratUHECR2022} and improving the dynamic range and the readout electronics, every WCD will also be equipped with a dual-polarized antenna measuring the radio signals from extensive air showers in the $30-80$\,MHz band. With the 1.5\,km spacing of the Auger surface detector (SD) array, this will allow the measurement of inclined air showers, with zenith angles beyond $\sim 65^\circ$, which have been both predicted by simulations \cite{HuegeUHECR2014} and confirmed with measurements of the Auger Engineering Radio Array (AERA) \cite{PierreAuger:2018pmw} to illuminate areas of tens of km$^2$ with detectable radio signals. For such geometries, the electromagnetic cascade of extensive air showers will be effectively absorbed in the atmosphere, and the WCDs will measure the virtually pure muon content of the particle showers. The radio antennas, in contrast, are purely sensitive to the electromagnetic cascade of the air showers. The combination of the two provides very good sensitivity to the mass composition of the primary cosmic rays, with different systematics from the combination of WCD and SSD measurements provided by AugerPrime for more vertical air showers.

In this article, we will shortly describe the concept and design of the \emph{Auger Radio Detector} (RD), report on its current status and plans for deployment, and finally present the results of an end-to-end simulation study of its expected performance.


\section{Concept of the Auger Radio Detector}

\begin{figure}[htb]
\centering
\includegraphics[width=0.5\textwidth,angle=270]{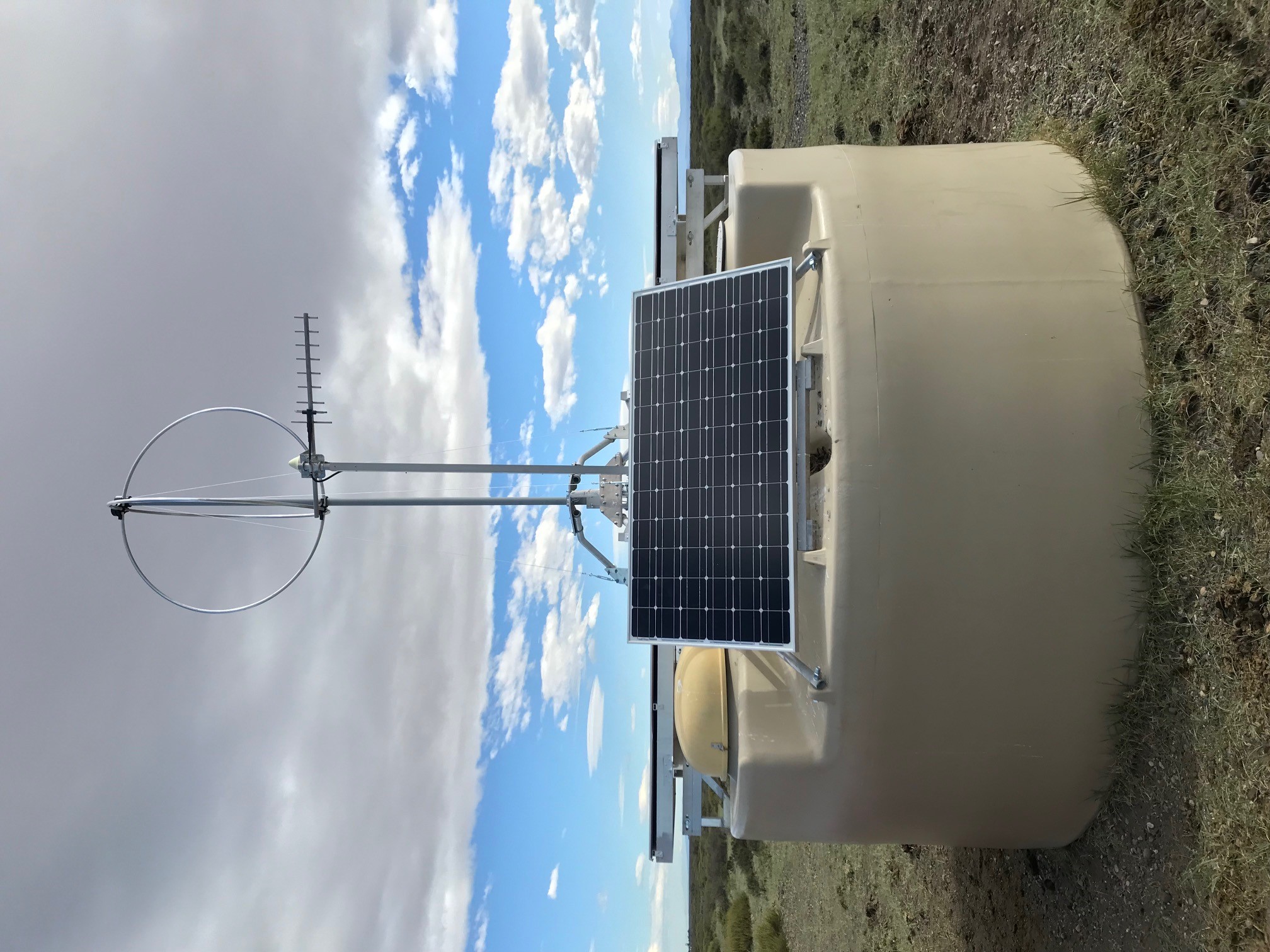}
\caption{Photo of a prototype RD station in the field. The dual-polarized SALLA antenna is mounted on a fiberglass mast held in place by an aluminum frame, which is connected directly to the WCD tank. Guy wires help suppress vibrations. The additional RD readout electronics is encapsulated in the dome visible in the foreground.}
\label{fig:rdstation}
\end{figure}

\begin{figure*}[htb]
\centering
\includegraphics[width=0.77\textwidth]{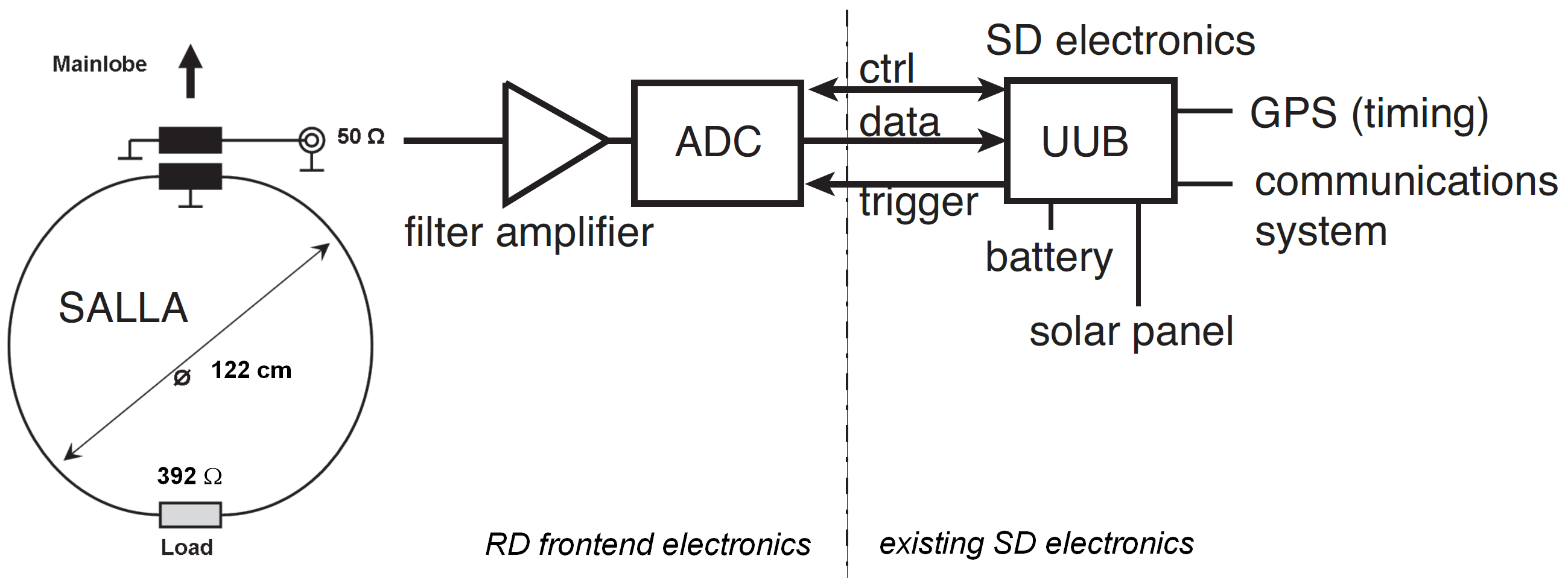}
\caption{Schematic diagram for the RD readout chain and its connection to the existing Surface Detector electronics via the Upgraded Unified Board (UUB). Updated from \cite{Horandel:2019hrm}.}
\label{fig:rdhardware}
\end{figure*}

The Auger RD builds on the available infrastructure of the upgraded SD. An upgraded detector station including a radio antenna is shown in Fig.\ \ref{fig:rdstation}. Two aluminum rings forming the dual-polarized antenna are mounted on a fiberglass mast in which the signal cables are contained. The mast is fixed by an aluminum frame directly connected to the structure of the SD tank (not touching the surface scintillator detector beneath). Guy wires additionally support the mast and reduce vibrations that could otherwise be induced by strong winds.

The antenna is a \emph{Short Aperiodic Loaded Loop Antenna} (SALLA) \cite{KroemerSALLA2009}, a revised and improved design based on the one originally developed in the context of AERA \cite{AERAAntennaPaper2012}. It features a simple mechanical design, minimizing cost and easing handling and maintenance. With its diameter of 122\,cm, it is tailored at the frequency range of interest of $30-80$\,MHz, for which it delivers a virtually uniform response with very little dispersion. The antenna features a 392\,$\Omega$ resistor at the bottom which shapes the antenna's main lobe towards the zenith and suppresses dependence on structures below the antenna, in particular the SSD, the WCD and potentially variable ground conditions. Although originally designed for AERA, the SALLA was not favored at the time because its resistor leads to a decreased sensitivity; only $\sim 10$\% of the captured signal intensity is available at the input of the low-noise amplifier at the top of the rings. However, this is not a relevant problem for the Auger RD, as it aims at the detection of the highest-energy cosmic rays, where signal strengths are not a limiting factor.

The concept for the readout is illustrated in Fig.\ \ref{fig:rdhardware}. Each of the two antenna channels is read out by an analog-to-digital converter with a sampling rate of 250\,MHz and a dynamic range of 12\,bits. Before digitization, the signals are amplified by a total of 36\,dB and filtered by a bandpass filter with a width of $30-80$\,MHz. An FPGA (not shown in the diagram) coordinates data exchange with the \emph{Upgraded Unified Board} (UUB) which queries 2,048 samples of the radio data whenever a trigger was received from the WCD. In the future, we plan to also include information from the radio detector in the trigger decision, which would be useful in particular for the detection of photon-induced air showers. 

Finally, the data are sent via the existing wireless communications system as part of the regular data stream. Further monitoring information, such as a regular characterisation of the Galactic radio background, will in addition be transmitted as part of a monitoring data stream.


\section{Status of the Auger Radio Detector}

The design of the RD hardware has undergone several iterations with various prototypes and has been finalized after completion of a critical design review. Since November 2019, a hexagonal configuration of seven prototype RD stations have been operated within an \emph{RD engineering array} (RDEA). The mechanical structures have proven to withstand the harsh conditions of the Argentinian pampa, and also the electronics have performed within specifications. At the site of the RDEA, the background situation is very favorable with very low rates of pulsed radio-frequency interference (RFI) and a comparatively clean measurement band.

\begin{figure*}[htb]
\centering
\includegraphics[width=0.5\textwidth]{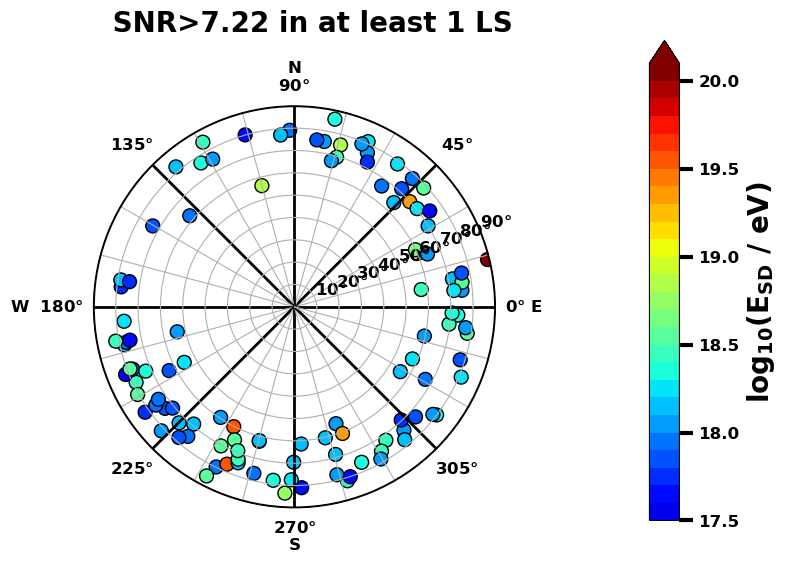}
\includegraphics[width=0.4\textwidth]{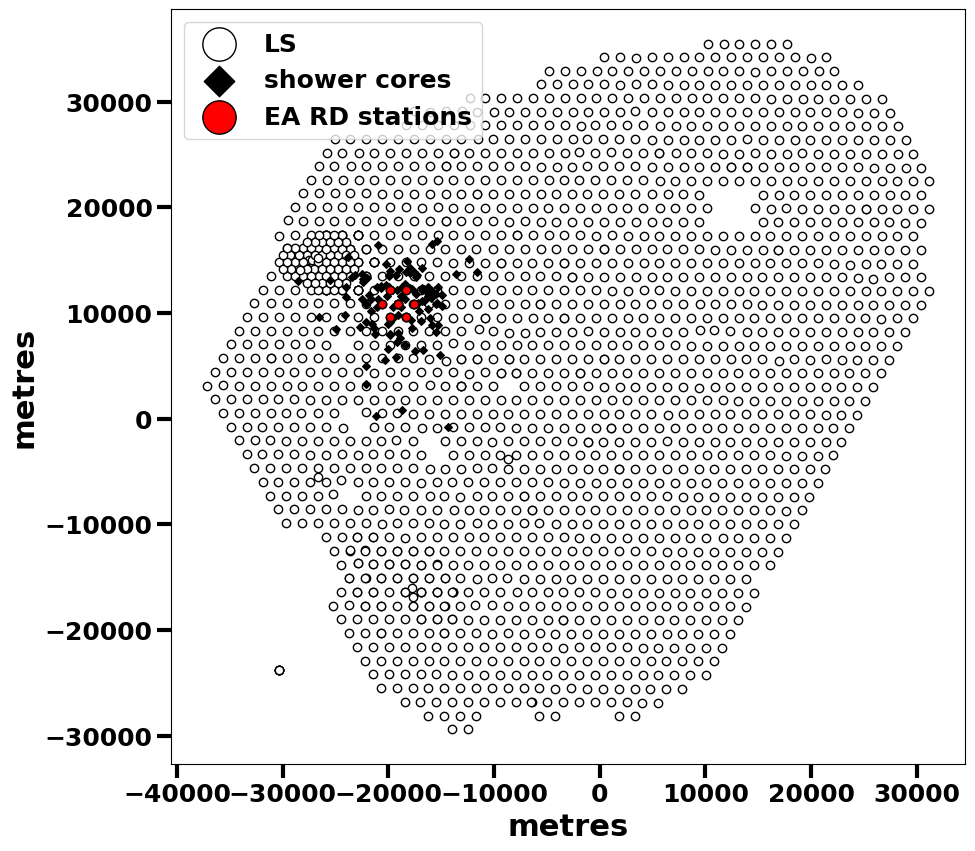}
\caption{Data taken with the RD engineering array over the course of 12 months. \emph{Left:} Angular distribution of air showers for which at least one radio pulse with a signal-to-noise ratio (SNR) exceeding a value of 7.22 (defined as the maximum amplitude squared divided by the RMS power in a noise window) was seen in at least one antenna at the expected pulse arrival time after triggering the readout by the WCDs. \emph{Right:} Distribution of the impact points of the showers for which an air-shower radio signal was detected.}
\label{fig:rdeaevents}
\end{figure*}

In Fig.\ \ref{fig:rdeaevents} we show events measured over the course of 12~months, triggered by the WCDs, for which at least one pulsed signal above Galactic background was seen at the expected time. The left panel shows the distribution of arrival directions, and as usual for air-shower radio emission, a north--south asymmetry can be seen due to threshold effects related to the angle the shower makes with the Earth's magnetic field (cf.\ magnitude of the Lorentz force). The right panel shows the distribution of the cores of the detected air showers.

For the calibration of the detector we perform calculations  of the antenna directional gain pattern using the NEC2 code and in-the-lab measurements of the responses of the LNAs, the filteramplifiers and the digitizers. These characteristics and full capability for reading in and processing measured data as well as simulations have already been included in the Offline analysis framework of the Pierre Auger collaboration \cite{ArgiroOffline2007}, building on the analysis functionality previously developed for AERA \cite{AbreuAgliettaAhn2011}. For an in-situ absolute calibration, we then compare the frequency- and local-sidereal-time-dependent power received from the sky with a model of the Galactic background folded through our above-mentioned detector response. The comparison shows that the two are in agreement within $\sim5$\% \cite{PierreAuger:2021bwp}, confirming the accuracy of the characterisation of the system and illustrating the feasibility of a continued absolute calibration on Galactic background signals \cite{Busken:2022mub}.

With the design finalized and proven, we have initiated mass production of the required 1,660 units plus spares in 2022. Except for some cables, the LNAs and the RD digitizer boards, the required hardware is already on site in Malargüe and being assembled for deployment. After a temperature-cycling and in-the-lab characterization of the LNAs, they and the missing cables will be shipped to Argentina shortly. The production of the digitizers currently suffers from delivery chain disruptions resulting in the unavailability of certain components. That said, deployment of a first series of complete detectors is envisaged in the first half of 2023, and the deployment of the full RD array is envisaged to be completed by the end of 2023.


\section{Expected performance of the Auger Radio Detector}

In an earlier publication \cite{2019ICRC_pont}, we had estimated the expected performance of the Auger Radio Detector in terms of achievable event statistics, detection threshold and potential for muon-number measurements on the basis of a signal model and an assumed background. Here, we present a study based on a fully realistic, end-to-end simulation of the actual detector response, Monte Carlo simulations with the CoREAS code \cite{Huege:2013vt}, and an event reconstruction approach tailored to radio detection of inclined air showers \cite{Schluter:2022mhq}, using realistic uncertainties and only information that will also be available in measured data. This allows us to provide a detailed account of the expected performance of the Auger RD in terms of event statistics, detection threshold, achievable energy resolution, potential for muon number measurements, and mass composition sensitivity.

\subsection{Simulation setup}

As a basis for this study, we use CORSIKA \cite{HeckKnappCapdevielle1998} simulations, including CoREAS \cite{Huege:2013vt} for the simulation of the radio emission, for proton, helium, nitrogen and iron nuclei arriving with zenith angles in the range from 65$^\circ$ to 85$^\circ$ following a sin$^2 \theta$ distribution, energies sampled uniformly in the logarithm of the energy in the range from 10$^{18.4}$\,eV to 10$^{20.2}$\,eV, and random, uniformly distributed azimuth angles. The atmosphere and magnetic field configuration were chosen with representative values for the site of the Auger Observatory. A total of $\sim$\,8,000 simulations each were generated with the QGSJETII-04 \cite{PhysRevD.83.014018} and Sibyll 2.3d \cite{Engel:2019dsg} interaction models and used to derive the results presented in this article. An additional 8,000 simulations probing the effects of more conservative particle thinning and the influence of seasonal changes in the atmospheric conditions were also generated and used for further cross-checks.

These simulations undergo a realistic end-to-end detector simulation using the Auger Offline analysis framework. The particle content of the showers is used to simulate the triggering of the WCD stations; only radio signals of particle-triggered stations are used in the following. The radio signals are convolved with the direction- and frequency-dependent antenna gain for the dual-polarized antennas, followed by bandpass filtering, signal amplification and digitization. Actual measured noise from the RDEA stations is added to the simulated ADC traces. A Gaussian smearing with a width of 5\% is applied to the ADC traces of individual antennas to mimic uncertainties and variations from antenna to antenna. Finally, the signal timing is smeared out with a Gaussian of width 6\,ns to mimic the accuracy reachable by the GPS timing system, yielding ``simulated traces like measured data''. 

\subsection{Detection efficiency, aperture and event statistics}

First, we study the detectability of air showers with the Auger RD. We deconvolve the detector response and reconstruct the electric field traces. On the basis of these we determine a signal-to-noise ratio (SNR), measured as the ratio of the square of the maximum, Hilbert-enveloped electric field, divided by the RMS power determined from a noise window. We consider a signal detectable if it exceeds an SNR of 10, which ensures a signal purity of better than 99\%. We count a shower as ``detected'' if  three antennas were triggered to be read out by the associated WCDs and exceeded this SNR-threshold. The detection efficiency we then calculate as the ratio of such-detected showers and the number of showers that triggered the SD. (Calculating the ratio to all simulated showers would underestimate the detection efficiency because of the muon number deficit in simulations; the SD trigger simulation would be too pessimistic.)

\begin{figure*}[htb]
\centering
\includegraphics[width=0.43\textwidth]{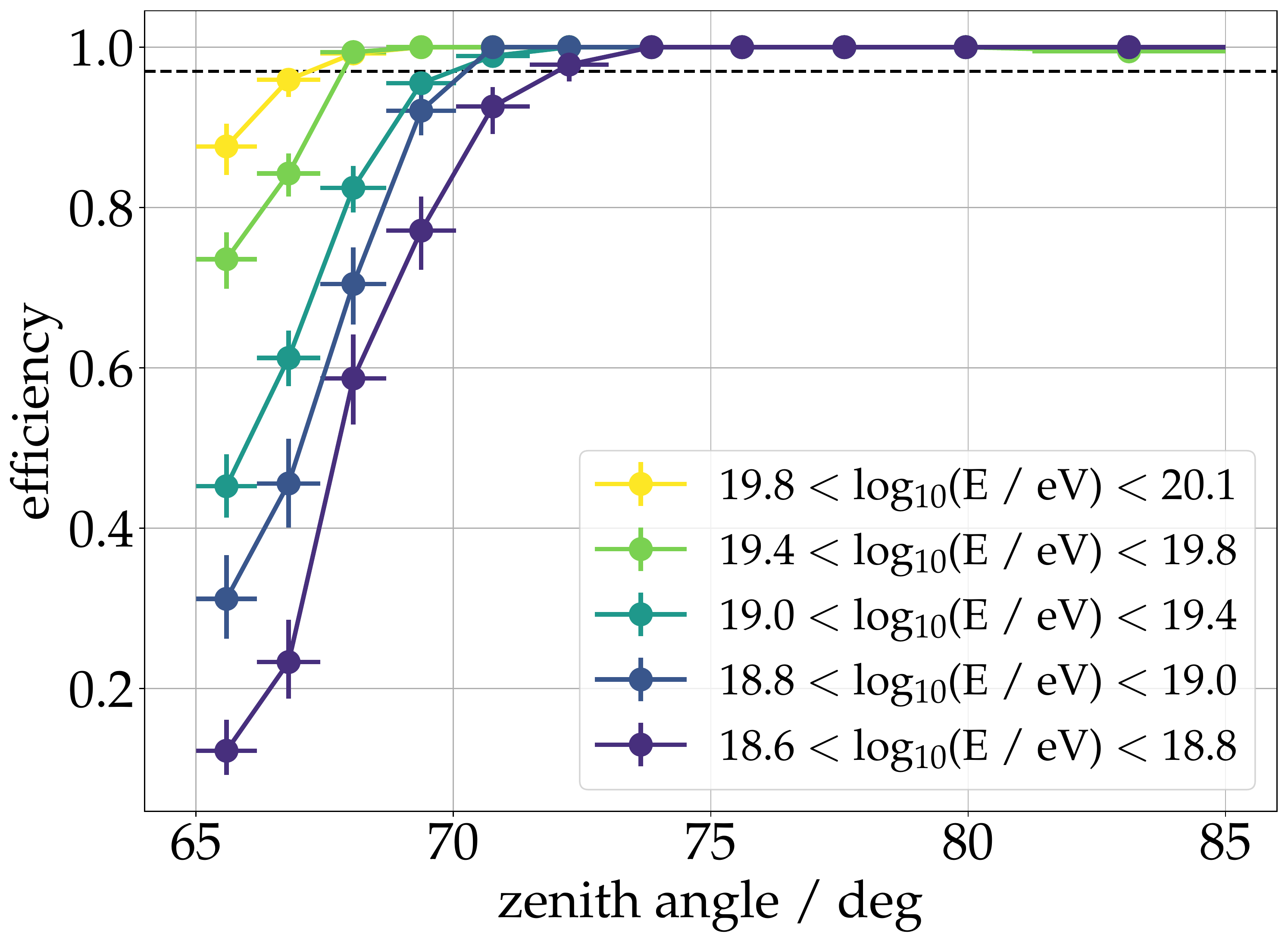}
\includegraphics[width=0.56\textwidth]{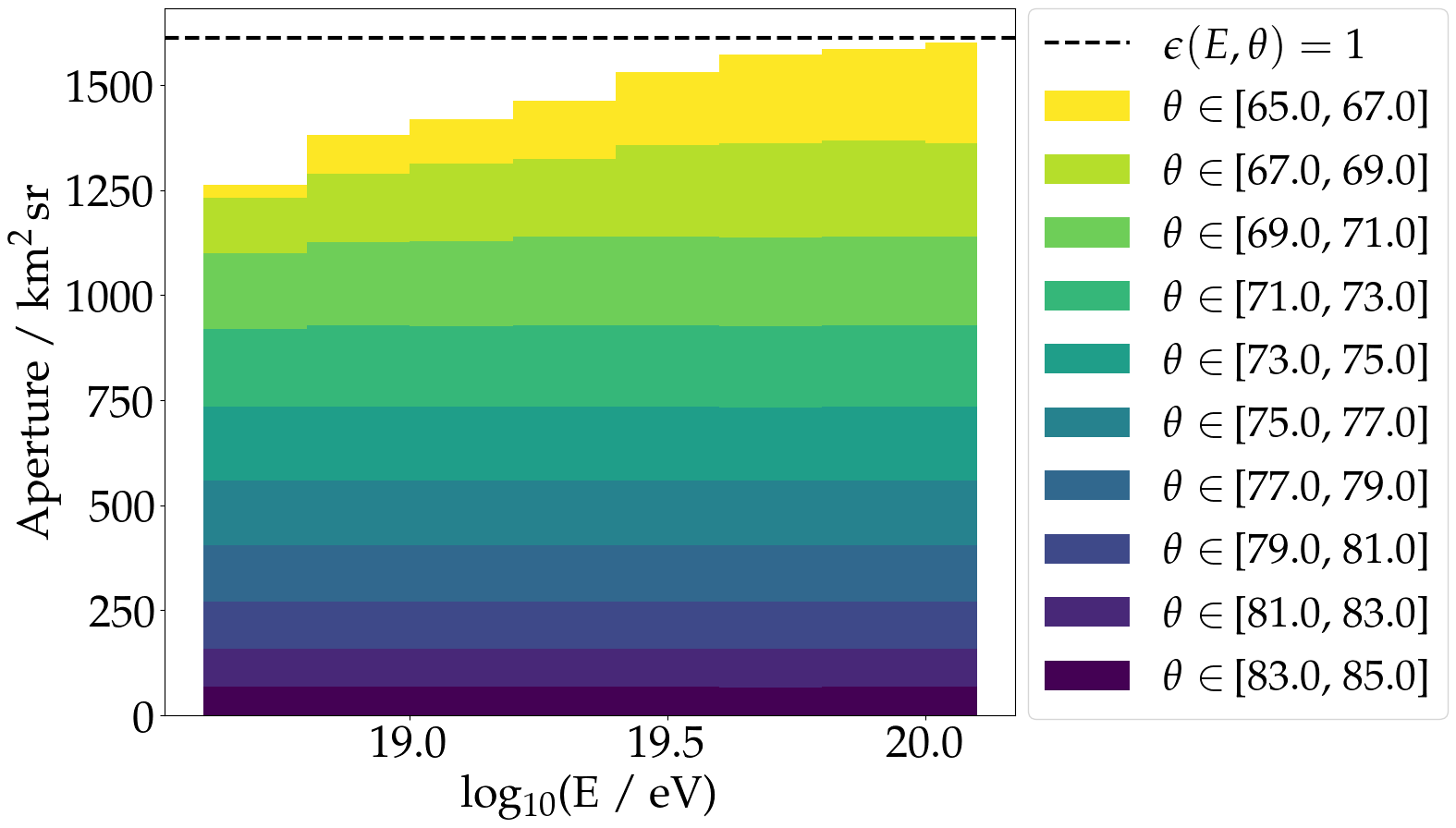}
\caption{Left: Detection efficiency with the Auger Radio Detector as a function of zenith angle an particle energy. The dashed line marks an efficiency of 97\%. The error bars denote statistical uncertainties. Right: Stacked histogram of the aperture of the Auger Radio Detector as a function of zenith angle and energy. This applies to \emph{contained} showers which have their impact point in the geometrical 3,000\,km$^2$ area of the Pierre Auger Observatory.}
\label{fig:effandapp}
\end{figure*}

The resulting detection efficiency with the Auger Radio Detector as a function of energy and zenith angle is shown in Fig.\ \ref{fig:effandapp} (left). It is obvious, and expected, that the zenith angle has a much higher influence on detectability than the energy, as it governs the size of the radio-emission footprint and thus is the main factor determining whether a signal can be detected in three antennas coincidentally. The Auger Radio detector will be fully efficient as of $\sim 70^\circ$ zenith angle.

Using the detection efficiency as a function of zenith angle and energy, we calculate the aperture of the Auger Radio Detector and show it as a stacked histogram in Fig.\ \ref{fig:effandapp} (right). Here, we only account for air showers which have their impact point inside the geometrical area of the Auger SD, i.e., only \emph{contained} events. A factor of $\cos \theta$ arises from the projection of the array area to the plane normal to the incoming direction and explains why the contributions of the most inclined air showers to the aperture become small, even though their detection is fully efficient for the full energy range from 10$^{18.6}$\,eV to 10$^{20.2}$\,eV. In contrast, the lower zenith angles provide a larger aperture at the highest energies, but detection becomes inefficient at low energies. Note that, for the highest energies, the aperture approaches the geometrical optimum. These results are in reasonable agreement with the earlier, much more simplified, estimates reported in reference \cite{2019ICRC_pont}.

\begin{figure*}[htb]
\centering
\includegraphics[width=0.78\textwidth]{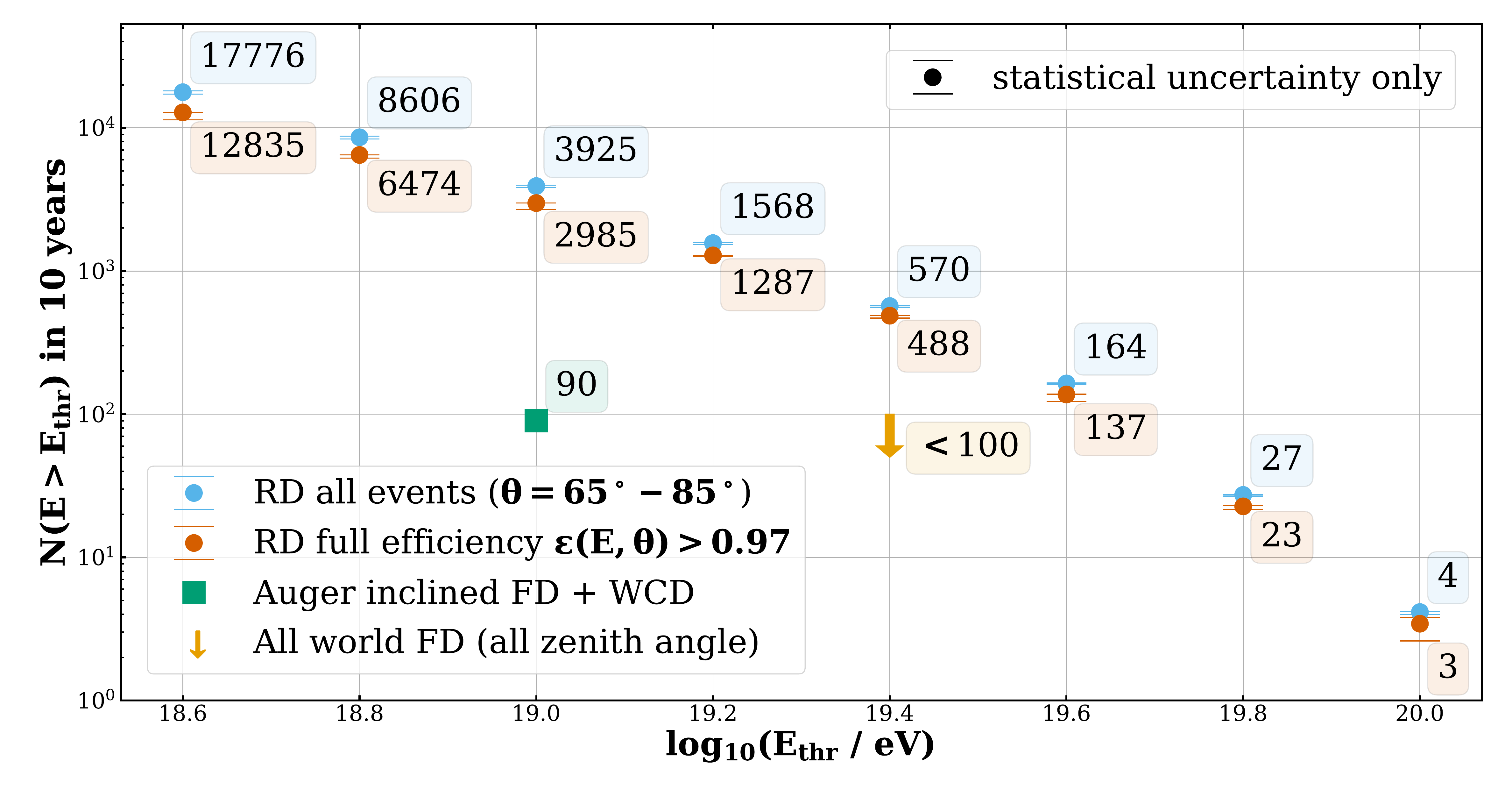}
\caption{Integral number of cosmic rays expected to be detected by the Auger Radio Detector for a measurement period of 10 years. Blue points denote all detected events, red points only those in bins of energy and zenith angle for which the detection efficiency is at least 97\%. For comparison, the number of inclined air showers measured with both the Auger Surface and Fluorescence Detectors is shown in green \cite{PierreAuger:2021qsd}, and the total statistics of air showers measured with Fluorescence Detectors worldwide at energies of 10$^{19.4}$\,eV or higher is shown in yellow \cite{Coleman:2022abf}.}
\label{fig:eventstats}
\end{figure*}

In a next step, we multiply the aperture with the cosmic-ray flux as published in reference \cite{PierreAuger:2020qqz} to determine the number of detectable events. For an assumed measurement period of 10~years, the expected event statistics are shown in Fig.\ \ref{fig:eventstats}. The quoted numbers denote the integral event statistics for showers above a given threshold energy. The blue data points refer to the expected event numbers if all detected events are counted; the red points refer to only those bins in zenith angle and energy for which the detection efficiency is at least 97\%. For the highest energies of $10^{19}$\,eV or higher, we expect between 3,000 and 4,000 cosmic rays in 10 years. 

\subsection{Expected energy resolution}

In a next step, we evaluate the resolution of the energy reconstruction achievable with the Auger Radio Detector. It should be noted that radio detectors measure the \emph{electromagnetic energy} of the air shower, i.e.,  the energy content of the electromagnetic cascade. This can be converted to cosmic-ray energy with knowledge of the \emph{invisible energy}, but strictly speaking is dependent on the mass of the primary particle. In terms of mass composition sensitivity, however, it is anyway preferable to directly use the electromagnetic energy in conjunction with the muon number as measured by the particle detectors, which will be explored in the following subsection.

\begin{figure*}
\centering
\includegraphics[width=0.51\textwidth]{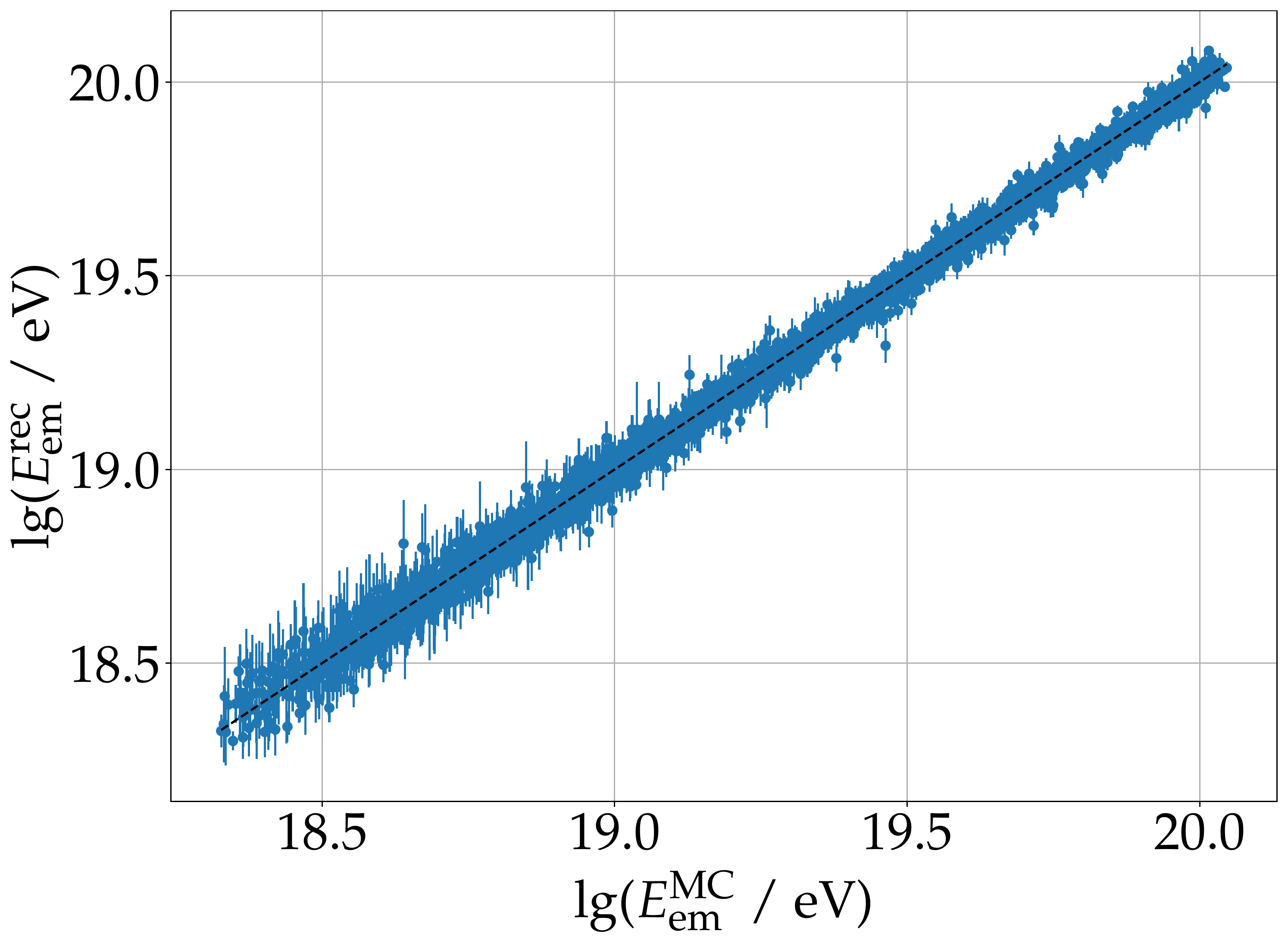}
\includegraphics[width=0.48\textwidth]{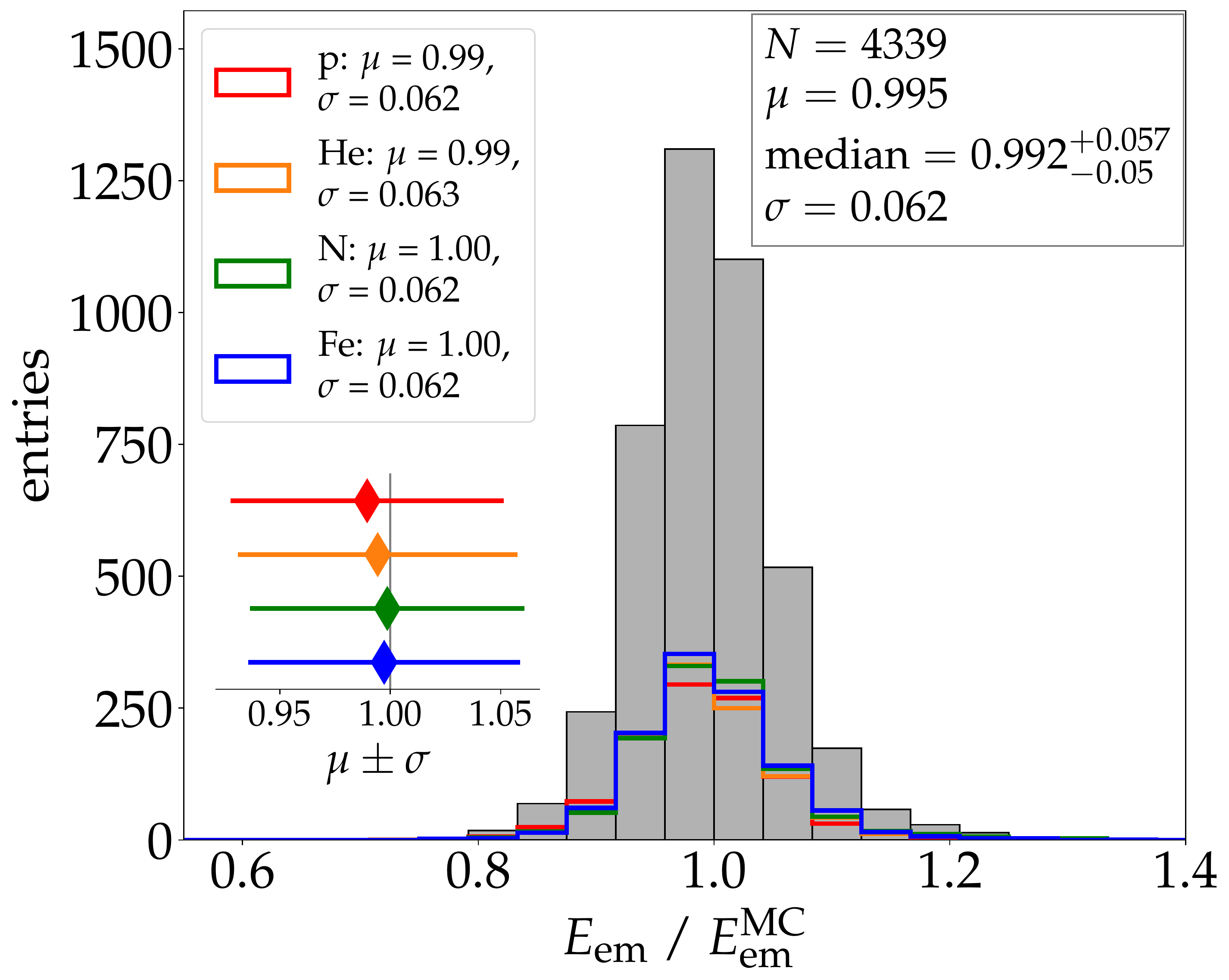}
\caption{Left: Reconstructed electromagnetic energy versus Monte Carlo true electromagnetic energy as expected for the Auger Radio Detector. Right: Histograms illustrating the performance of the energy reconstruction for the whole simulation set as well as the subsets for different primary particle types.}
\label{fig:energyreco}
\end{figure*}

To reconstruct the electromagnetic energy \Eem from radio measurements, one first determines the energy fluence, i.e., the energy deposited in the form of radio emission in the $30-80$\,MHz band per unit area, at every antenna position with a detectable signal. After an area integration and the application of corrections for the effects of the angle to the magnetic field and the air density at shower maximum, one can then determine the \emph{corrected radiation energy}, an accurate estimator for the energy in the electromagnetic cascade of an air shower \cite{AERAEnergyPRD,AERAEnergyPRL,Glaser:2016qso}.

For the Auger Radio Detector, we use a reconstruction procedure tailored specifically to inclined air showers based on a signal model that explicitly separates the geomagnetic and charge-excess emission in inclined air showers and also takes care of geometrical early-late asymmetries in the signal distributions on ground \cite{Schluter:2022mhq}. In the following, we only use air showers with a detectable signal in at least five antennas, and we discard events with zenith angles below 68$^\circ$. We estimate an initial impact point of the shower from the SD reconstruction and the arrival direction of the shower from RD signals using a spherical wavefront model. With some loose quality cuts on the sampling of the radio-emission footprint and the quality of the signal-distribution fit, survived by $\sim 95$\% of the simulated showers, we retrieve the energy reconstruction performance illustrated in Fig.\ \ref{fig:energyreco} (left). The scatter plot illustrates a 1:1 correlation between reconstructed and true electromagnetic energy, with an improvement of the resolution towards higher energies (and thus signal-to-noise ratios). In the right panel, a more quantitative characterisation of the reconstruction quality is shown in terms of histograms of the ratio of reconstructed versus true electromagnetic energy, both for the simulation set as a whole and for the subsets for particular types of the primary particle. No significant biases are present; the energy resolution is at the level of $\sim 6$\%.

\subsection{Expected mass composition sensitivity}

The goal of the upgraded Pierre Auger Observatory is to exploit separate measurements of the electromagnetic and muonic content of air showers for mass composition measurements. (The combined measurement breaks the degeneracy in the muon number scaling with both energy and mass of the primary particle.) While for vertical air showers the separate measurement of the two components is achieved by the combination of the WCD and the SSD, for inclined air showers the combination of the RD and WCD is used. The RD delivers an accurate measurement of the electromagnetic energy of an air shower, while the WCD measures the virtually pure muon content of inclined air showers. The combination thus allows us to study the muon content of extensive air showers, as has previously been performed using the combination of WCD and Fluorescence Detector data \cite{PierreAuger:2021qsd}, yet with orders of magnitude lower statistics than will be possible with the RD, as illustrated already in Fig.\ \ref{fig:eventstats}.

\begin{figure*}
\centering
\includegraphics[width=0.49\textwidth]{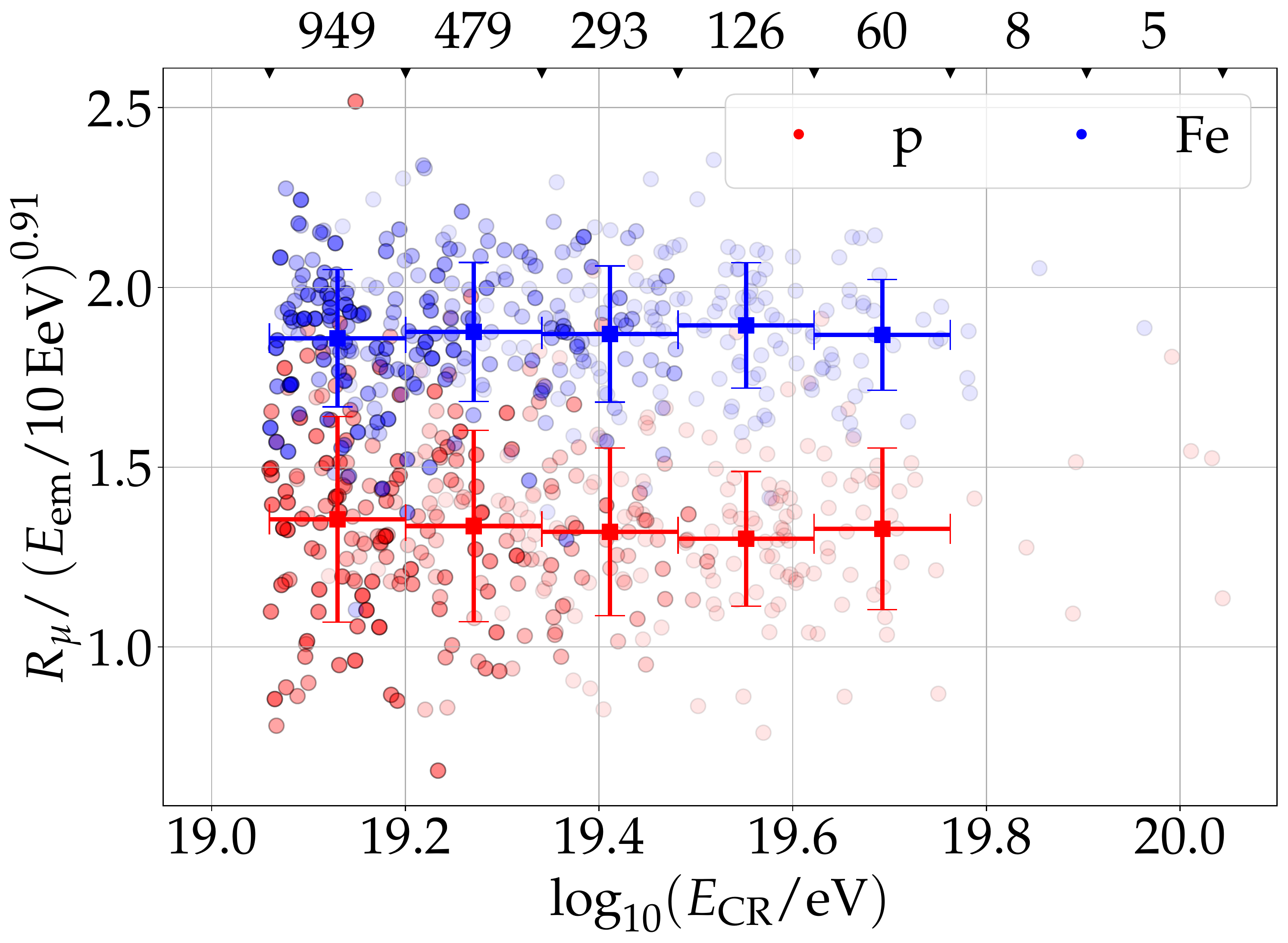}
\includegraphics[width=0.49\textwidth]{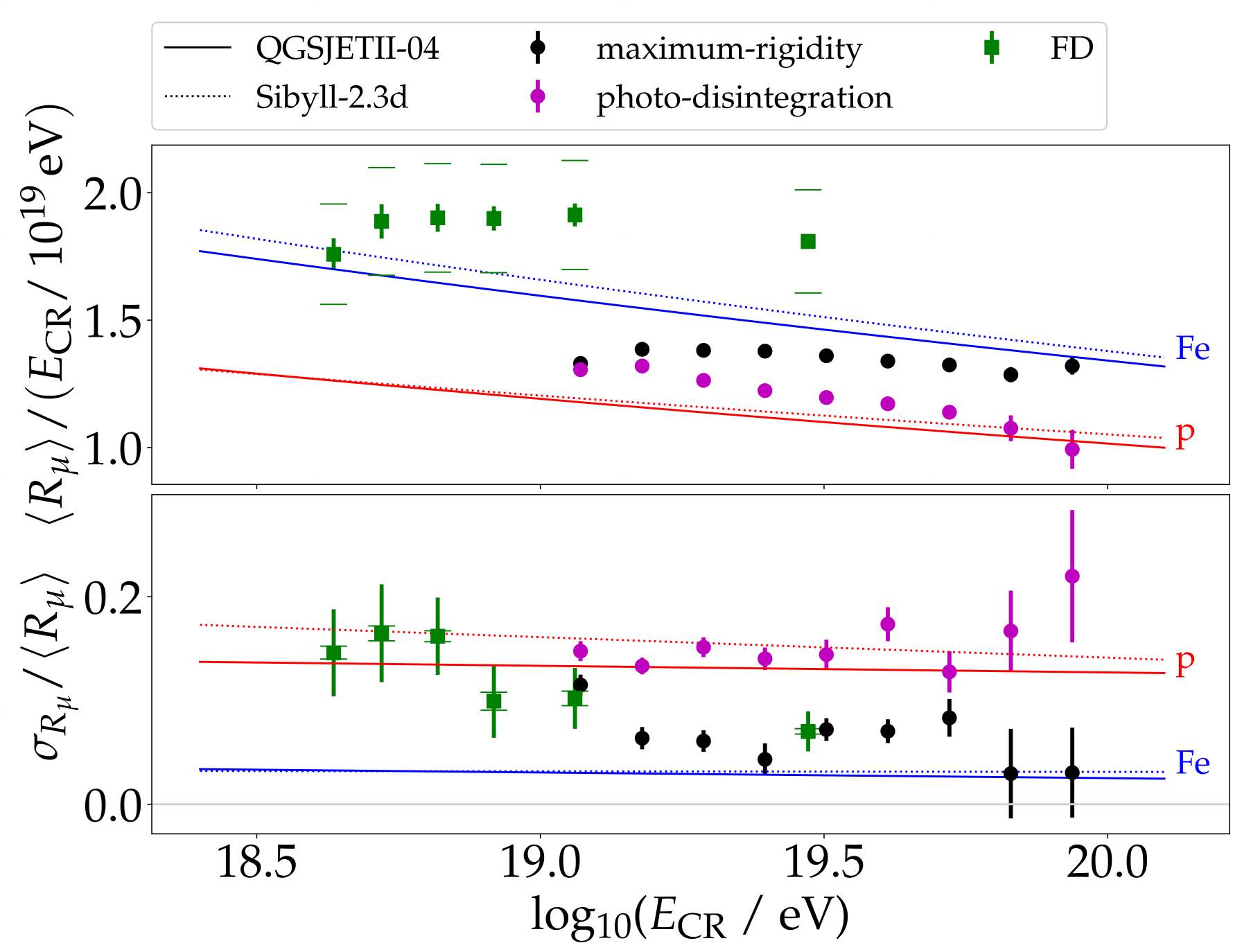}
\caption{Left: Separation power between proton and iron primaries of the relative muon number normalized by the electromagnetic energy as a function of cosmic-ray energy. Right: Discrimination power between to different mass composition scenarios using the mean muon number (top) and the fluctuations in the number of muons (bottom) for combined measurements with the Auger Radio and Water Cherenkov Detectors in 10 years. In green, for comparison, measurements made previously with the Auger Fluorescence and Water Cherenkov Detectors. }
\label{fig:mass}
\end{figure*}

We first investigate an (unrealistic) mixture of 50\% proton and 50\% iron showers that have been reweighted to follow the actual, steep energy spectrum of cosmic rays at reconstructed electromagnetic energies above $10^{19}$\,eV with the statistics expected for ten years of measurement time. With the standard SD reconstruction for inclined air showers, we determine $R_\mu$, a measure for the muon content in air showers relative to the expectation for a simulation template at an energy of $10^{19}$\,eV. We remove a non-linear scaling with electromagnetic energy and define $r = \Rmu / (\Eem / 10\,\mathrm{EeV})^{0.91}$. The scatter plot and corresponding profiles for $r$ shown in Fig.\ \ref{fig:mass} (left) illustrate the separation between proton and iron primaries in the normalized relative muon number. The figure of merit $\mathrm{FOM} = |\mean{r_\mathrm{p}} - \mean{r_\mathrm{Fe}}| / \sqrt{\sigma_{r_\mathrm{p}}^2 + \sigma_{r_\mathrm{Fe}}^2} = 1.60 \pm 0.05$, which is an excellent value comparable to that of depth of shower maximum measurements with a near-perfect resolution below 10\,g\,cm$^{-2}$. Key to this success is the excellent energy resolution of $\sim 6$\% of the Auger Radio detector; the sensitivity would strongly suffer from a worse energy resolution, especially for the steep spectrum of cosmic rays at the highest energies.

Finally, we investigate the mass composition sensitivity for more realistic mixed composition scenarios, namely the ones chosen for the original preliminary design report for the upgrade of the Pierre Auger Observatory  \cite{PierreAuger:2016qzd}. One of the scenarios is a maximum-rigidity scenario whereas the other one represents a mass composition as expected from photo-disintegration. Again, we use the actual, steep cosmic ray spectrum and the expected event statistics for ten years of measurements. The results are shown in Fig.\ \ref{fig:mass}. Note that we here normalize on cosmic-ray energy \Ecr after converting to it from \Eem with a linear function determined from simulations, $\Ecr = \Eem \cdot \left[1.1426 - 0.0328 \log_{10}(\Eem / 10\,\mathrm{EeV})\right]$. The upper panel shows the mean number of muons as a function of energy. The error bars show the statistical uncertainties and illustrate the vastly improved statistics with respect to previous measurements with the Fluorescence and Water Cherenkov Detectors \cite{PierreAuger:2021qsd} shown by the green data points. However, measurements of the mean number of muons suffer from relatively large systematic uncertainties, as illustrated by the error caps of the green data points, arising to a large extent from uncertainties in the absolute energy scale of the measurement. The combined RD/WCD data are expected to have a systematic uncertainty on a similar scale as shown here for the FD/WCD data. The situation is much more favorable when measuring the \emph{fluctuations} in the number of muons as a function of primary energy, as illustrated in the lower panel. Here, systematic uncertainties due to the absolute energy scale cancel out. Measurements with the Radio and Water Cherenkov Detectors will therefore provide a powerful means to differentiate scenarios for the mass composition of cosmic rays at energies above $10^{19}$\,eV.


\section{Conclusions}

We have reported on the design, status, and expected performance of the Auger Radio Detector. This detector will consist of a dual-polarized short aperiodic loaded loop antenna on top of each surface detector, measuring radio signals in the $30-80$\,MHz band. The design of the Radio Detector is final and has been proven to work in an engineering array of seven detectors operating since November 2019. Measurements with the engineering array have successfully confirmed the compatibility of the calibration in the laboratory with an absolute calibration on the Galactic radio emission, and have also successfully demonstrated detection of radio pulses from extensive air showers. Mass production of the RD components is mostly complete, except for the digitizer electronics which suffer from continued delivery chain disruptions. Nevertheless, complete deployment of the Auger Radio Detector is envisaged for the year of 2023.

Meanwhile, the Offline analysis framework of the Pierre Auger Observatory has successfully been enabled for analyzing RD data. The detector has been characterized in detail in the analysis software, allowing us to perform an end-to-end simulation study of the expected performance of the RD on the basis of CoREAS simulations and measured background data. With this, we have shown that the RD will achieve full efficiency for air showers above $\sim 10^{18.6}$\,eV for zenith angles above $\sim 70^\circ$. We have quantified the aperture and predicted the expected number of detectable cosmic rays as a function of energy, yielding between 3,000 and 4,000 events in 10 years above 10$^{19}$\,eV.

Furthermore, an algorithm for the reconstruction of the electromagnetic energy of air showers measured with the RD has been developed and implemented in the Offline analysis framework. Under fully realistic conditions, employing realistic uncertainties and using only information that will also be available in actual measurements, this reconstruction achieves a resolution on the electromagnetic energy of $\sim 6$\%, with no significant bias, not even between different types of primary particles.

This excellent energy resolution is key to exploiting the mass composition sensitivity of combined measurements of the Radio and Water Cherenkov Detectors. We have studied this sensitivity for a realistic, steep spectrum of air showers, showing that the figure of merit for a separation of proton and iron primaries lies at an excellent value of 1.60. For more realistic mass composition scenarios, we have demonstrated that in particular the fluctuations in the number of muons measured by the combination of Radio and Water Cherenkov Detectors will be a powerful means to determine the mass composition of cosmic rays at the highest energies.

\section*{Acknowledgements}
We are very grateful to all the agencies and organizations that support the Pierre Auger Observatory (https://www.auger.org/collaboration/funding-agencies). For this work in particular, we acknowledge the support from the European Research Council Advanced Grant \#787622 and the Dutch Research Council.

\end{document}